\documentstyle[12pt,epsfig]{article}

\def\beq{\begin{equation}}
\def\eeq{\end{equation}}
\def\d{\delta}

\def\g{\gamma}

\def\e{\epsilon}

\begin{document}
\title{\bf Cosmological CMBR dipole in open universes~?}
\author{David Langlois \\ 
\it D\'epartement d'Astrophysique Relativiste et de Cosmologie,\\
\it UPR 176, Centre National de la Recherche Scientifique, \\
\it Observatoire de Paris, 92195 Meudon, France }
\date{\today}
\maketitle
 
\par\bigskip

\begin{abstract}
The observed CMBR dipole is generally interpreted as a Doppler effect 
arising from the motion of the Earth relative to the CMBR frame. 
An alternative interpretation, proposed in the last years,
 is that the dipole results from  
ultra-large scale isocurvature perturbations. We examine this idea in the 
context of open cosmologies and show that the isocurvature interpretation
is {\it not valid} in an open universe, unless it is extremely close to 
a flat universe, $|\Omega_0 -1|< 10^{-4}$.
\end{abstract}

\section{Introduction}
In the standard Big Bang scenario, the oldest relic from the early 
universe is the Cosmic Microwave Background Radiation (CMBR). This 
radiation is remarkably close to isotropy. When it is decomposed into 
multipoles,
\begin{equation} 
{\Delta T\over T}(\theta, \phi)=
\sum_{l=1}^{\infty}\sum_{m=-l}^{l} a_{lm}Y_{lm}(\theta,\phi), \label{fluct}
\end{equation}
 one finds that the dipole contribution is of the order 
$(\Delta T/ T)_{l=1}\sim 10^{-3}$, while the contribution from higher 
multipoles is only $(\Delta T/ T)_{l>1}\sim 10^{-5}$ (see e.g.\cite{kogutetal}).
Usually, the dipole term is interpreted as a Doppler effect, i.e. as the 
consequence of our local motion with respect to the ``CMBR rest frame'', whereas 
the other multipoles are  accounted for by primordial cosmological 
perturbations. However, one cannot reject {\it a priori} the possibility
that a significant part of the dipole originates as well from cosmological 
perturbations. In fact, this alternative idea was stimulated by the 
results of Lauer and Postman \cite {lap} in 1991, who found a dipole 
in distant Abell clusters inconsistent with the CMBR dipole. Other 
observations based on nearby galaxies, IRAS galaxies or  distant supernovae 
tend on the other hand to favour the orthodox interpretation, but are still
inconclusive. 

Let us be more precise on the distinction between a 
cosmological dipole and a local dipole. If our universe was perfectly homogeneous
and isotropic then there would be no cosmological dipole; a comoving 
observer would not detect any dipole whereas a non comoving observer 
would measure a dipole, in this case a purely local 
dipole, due to his motion. In a perturbed model, it
is more delicate to distinguish between a local and a cosmological dipole.
The perturbation of any quantity can be decomposed into 
contributions from different scales. The cosmological dipole is then 
due to very large scale perturbations, typically scales of the order 
of the Hubble radius, and the local dipole is due to the small scale 
component (the dominant one) of the observer peculiar velocity.

The purpose of this paper is to explore the origin of the dipole from a theoretical 
point of view and to ask whether it is possible, in the context of 
 Friedman-Lemaitre-Robertson-Walker (FLRW) cosmologies 
with gaussian random fields of linear perturbations, to 
obtain a {\it cosmological} dipole and higher multipoles compatible with the 
observations. To discriminate between a local Doppler dipole and  a cosmological dipole is more, in our view, than a mere academic exercise because the Doppler 
assumption enters in the quadrupole analysis of the observations. The 
reason is that our motion would not only induce a dipole but also a
quadrupole, and other multipoles.
 
It was shown by Paczynski and Piran \cite{pp} that a strong dipole can be 
found within a particular Tolman-Bondi model with an inhomogeneous 
radiative component. In fact, as was shown recently by Langlois and Piran 
\cite{lp}, this result can be obtained, more generally,
 in a perturbed FLRW {\it flat} cosmology, from {\it ultra large} scale 
(larger than the current Hubble radius) {\it isocurvature} perturbations.
Here, we reinvestigate this possibility in the context of {\it open}
cosmologies. 

The influence of stochastic cosmological perturbations with wavelengths
larger than the current Hubble radius on the CMBR temperature 
fluctuations was examined initially by Grishchuk and Zel'dovich \cite{gz}. 
Their analysis was restricted to  perturbations in a flat background
for which the Fourier expansion is available. When one considers an open 
universe one needs the complicated formalism of mode functions in an
hyperbolic space. Recently, Lyth and Woszczyna \cite{lw} stressed 
the importance of supercurvature modes, which had been ignored previously,
 when dealing with random fields. With this new element, the influence of 
ultra large scale perturbations on the CMBR for open universes was then 
investigated in references \cite{lw} and \cite{gllw}.
The present work extends these papers in two respects. First, we focus 
our attention on the dipole, which was not treated previously, and, second,
we allow not only for adiabatic perturbations but also for isocurvature 
perturbations.

The paper is organized as follows.
In section 2,  we recall the main steps of the derivation of the Sachs-Wolfe effect, responsible of the large angle
anisotropies in the CMBR. Our approach follows the
work of Panek \cite{panek}, although our notation is different. 
The final result is valid for  flat or curved FLRW background geometries and
is expressed in terms of gauge-invariant variables.
Section 3 gives the CMBR multipoles in  a flat geometry. Although these
results were already given in \cite{lp}, the actual presentation is
 original  and provides a bridge to the formalism needed for an open cosmology.
In section 4, the case of an open cosmology is considered. Our notation 
follows essentially Refs \cite{lw} and \cite{gllw}.
Finally, section 5 contains the conclusions of this work.

\section{The Sachs-Wolfe effect}

Our background geometry is described by a FLRW metric,
\beq
ds^2=a^2(\eta)\left(-d\eta^2+\gamma_{ij}dx^idx^j \right),
\eeq
where $dl^2=\gamma_{ij}dx^i dx^j$ is the metric of a flat or curved 
maximally symmetric space (a flat space corresponds to $K=0$ and
an open space to $K=-1$, where $K$ is the normalized space curvature). 
The CMBR is composed of cosmological photons, which became free 
after the recombination of protons and electrons and the subsequent 
decoupling of matter and radiation at a redshift $z_{ls}$. For numerical 
applications we shall take $z_{ls}=1000$ ($z_{ls}$ depends in fact 
on the spatial curvature but this dependence is very weak \cite{kt} and can be
 safely ignored here).
All the photons reaching us now were emitted on a physical surface at the 
epoch of decoupling, called the last scattering surface, and   
defined physically by $n_e=const$, where
$n_e$ is the density of free electrons. The universe will be  supposed to 
be matter dominated from the last scattering time to now, which is an excellent 
approximation except for low values of $\Omega_0$.

Now, we consider a perturbed FLRW universe, endowed with the metric (in the 
longitudinal gauge)
\beq
ds^2=a^2(\eta)\left[-(1+2\Phi)d\eta^2+(1-2\Psi)\gamma_{ij} dx^i dx^j\right].
\label{metric}\eeq
Only perturbations of the scalar type will be considered here (see e.g. 
\cite{bardeen}) and, for simplicity, the anisotropic stress of the matter 
will be  supposed to vanish so that the Einstein equations yield
\beq
\Phi=\Psi.
\eeq
 Because of the presence of geometric perturbations, 
photons will be redshifted with slight differences depending on  their 
position of emission on the last scattering surface. The resulting 
fluctuations in  the temperature measured by an observer were first 
calculated by Sachs and Wolfe \cite{sw}. The origin of this effect 
is purely geometric and is dominant only for large angular scales, which 
correspond essentially to scales larger than the Hubble radius at the 
epoch of last scattering. On smaller scales, causal processes must be 
taken into account. In this work, we are interested only by the first 
multipoles of the CMBR, which are dominated by the last scattering 
superhorizon modes, so that only the Sachs-Wolfe effect is relevant.
In the following, the Sachs-Wolfe effect is rederived shortly following
the more modern calculation of Panek \cite{panek}. However our notations are 
different and the final result is expressed in another form.

  Denoting $k^\mu$ the vector tangent to a null geodesic,
 the evolution of the photon is governed by the geodesic equation:
\beq
k^\nu\partial_\nu k^\mu+\Gamma^\mu_{\sigma\tau}k^\sigma k^\tau=0,
\eeq 
where the $\Gamma^\mu_{\sigma\tau}$ are the Christoffel symbols.
We write the perturbed tangent vector $k^\mu=\{\nu(1-M), -\nu e^i+ P^i\}$, 
where the terms in $\nu$ represent the unperturbed solution; $e^i$ 
is a unit spatial vector, i.e. such that $\gamma_{ij}e^i e^j=1$; $M$ is the 
perturbation of the frequency, $P^i$ the perturbation of the spatial 
direction. Here, we are interested only in the frequency perturbation and
 thus keep only the time component of the geodesic equation. After
using  the identity $g_{\mu\nu}k^\mu k^\nu=0$
at first order, which reads
\beq
\gamma_{ij}e^iP^j=\left(M-\Phi-\Psi\right)\nu,
\eeq
in order to eliminate the  $P^i$ in the time component of the geodesic equation
 at first order, one finds eventually
\beq
{d M\over d\lambda}\equiv (\partial_\eta-e^i
\partial_i)M=\Phi'- 2 e^i\partial_i\Phi - \Psi'.
\label{geodesic}
\eeq
The prime stands for a derivative with respect to the 
conformal time $\eta$.
Note that the  parameter $\lambda$ is related to the affine parameter $\tau$
of the unperturbed geodesic by the relation
${d\tau/d\lambda}=1/\nu$. The prime stands for derivative with respect to the 
conformal time $\eta$. Equation (\ref{geodesic}) corresponds to Eq. (29)
of \cite{panek}. 

Typically one can model the content of the universe from  the epoch of the 
last scattering to now by two fluids, a pressureless baryon 
component and a radiation component, described locally by a black-body 
spectrum.
Let us define the temperature of the photons as seen by observers moving
with the baryon component, so that the temperature ratio between emission 
and reception is given by 
\beq
{T_R\over T_E}={(k^\mu u^b_\mu)_R\over (k^\mu u^b_\mu)_E}.
\eeq
In the perturbed geometry (\ref{metric}), one thus obtains
\beq 
{T_R\over T_E}={a_E\over a_R}\left\{1+\left[\Phi-M+e^i\partial_i V\right]^R_E
\right\},
\eeq
where the velocity potential $V$ is defined by $\d u_i^b\equiv a \partial_iV$.
As a consequence, the temperature fluctuations on the sky measured 
by an observer today (denoted by the subscript $0$)  are given by
\beq
\left({\d T\over T}\right)_0(\theta,\phi)=\left({\d T\over T}\right)_e +
\left({\d a\over a}\right)_e+\left[\Phi-M+e^i\partial_i V\right]^0_e.
\label{temp}
\eeq
The left hand side term is a function of the celestial coordinates 
corresponding to the direction of observation for the observer.
 The right hand side is expressed 
as a function of  the   emission point, of coordinates $(\eta_e,x^i_e)$, 
defined as the intersection of the last scattering surface with 
the  null geodesic going through the observer with the given direction , 
as well as
the observer position  $(\eta_0,x^i_0)$. In general the 
physical last scattering surface is distinct from the constant time
hypersurface $\eta=\eta_{ls}$, but one can replace the first two terms on the 
right hand side of (\ref{temp})
by $(\d T/T)(\eta_{ls},x^i_{ls}\simeq x^i_e)$ (i.e. as a function of the intersection 
point between the light geodesic and the hypersurface $\eta=\eta_{ls}$).
This follows from  the local conservation 
law $aT=const$ for free radiation. Then, expressing the  term 
$\left[\Phi-M\right]_e^0$ in (\ref{temp}) 
 as an integral over the null geodesic and using equation (\ref{geodesic}),
one finds 
\beq
\left({\d T\over T}\right)_R={1\over 4}\d_{\gamma| ls}+(h V)_{ls}
+\left[-\Phi+e^i\partial_i V\right]_{ls}^0 
+\int_{ls}^0 d\lambda \left(\Phi'+\Psi'\right),
\label{sw}
\eeq
where $h\equiv a'/a$. Using the Stefan law $\rho_\g\propto T^4$, we have 
replaced the temperature fluctuations by fluctuations of the radiation
energy density and also introduced  the comoving energy density perturbation 
$\d$, which can be derived from the 
energy density perturbation in the longitudinal gauge by the expression 
$\d= (\d\rho/\rho)_L+(\rho'/\rho)V$. To obtain  (\ref{sw}), we
have used implicitly
the equality of the velocities of radiation and baryonic matter at the time
of last scattering, which results from the preexisting tight coupling between 
the two fluids.
The   term  $\d_\g/4$ in (\ref{sw}) represents the intrinsic temperature fluctuations 
on the last scattering surface. The rest of the expression 
is the Sachs-Wolfe effect. In (\ref{sw}) all the quantities at the 
last scattering epoch are now evaluated on the hypersurface $\eta_{ls}$
which does not necessarily coincides with the physical last scattering 
surface (but all the corrections would be second order perturbations).

It is instructive to decompose the Sachs-Wolfe expression into 
several components and to rewrite (\ref{sw}), after dropping 
the term $-\Phi_0$ which contributes only to the monopole,
 in the form 
\beq
\left({\d T\over T}\right)_R=\left({\d T\over T}\right)_{int}
+\left({\d T\over T}\right)_{pSW}+\left({\d T\over T}\right)_{Dop}
+\left({\d T\over T}\right)_{ISW},
\eeq
with 
\beq
\left({\d T\over T}\right)_{int}={1\over 4}\d_{\gamma| ls}, \quad
\left({\d T\over T}\right)_{pSW}=(\Phi+hV)_{ls}, \quad
\left({\d T\over T}\right)_{Dop}=e^i\partial_i(V_0-V_{ls}), 
\label{decompos1}
\eeq
and
\beq
\left({\d T\over T}\right)_{ISW}=\int_{ls}^0 d\lambda \left(\Phi'+\Psi'\right).
\eeq
The term $(\Phi+hV)_{ls}$ will be
called the proper (or ordinary) Sachs-Wolfe effect (pSW), because only this 
term was computed in the pioneering paper of Sachs and Wolfe \cite{sw}.
The term $e^i\partial_i(V_0-V_{ls})$ is the difference between the 
observer velocity and the emission velocity along the line of sight and
corresponds to a Doppler effect (Dop). Note that the local dipole is due 
only to a term of the form $e^i v_i^0$, where $v_i^0$ is the contribution 
of small scales to our peculiar velocity; this contribution, which comes
from the nonlinear evolution of the perturbations,  will be 
ignored in the rest of the paper where we consider only the effect 
of very large scales.  Finally the integral term is 
usually called the integrated Sachs-Wolfe effect (ISW).

The linearized Einstein equations and fluid equations (see e.g. \cite{mfb} or 
\cite{ks}) 
can then be used to express all the 
terms in (\ref{sw}) in terms of only one quantity, the most convenient 
being the gravitational potential $\Phi$. For example, 
 the  velocity potential $V$ can be  related to  $\Phi$  via  the Euler equation and the conservation 
equation:
\beq
V=-{2\over 3(h^2+K)}(\Phi'+h\Phi).\label{v}
\eeq
Via the Einstein equations, one gets a relativistic generalization of Poisson 
equation relating the gravitational potential to the total comoving
energy density $\d_T$, which reads: 
\beq
(\triangle+3K)\Phi={3\over 2}(h^2+K)\d_T. \label{poisson}
\eeq
At this stage one must distinguish two kinds of perturbations:
adiabatic and isocurvature perturbations. Any general perturbation
can be seen as a linear combination of these two kinds.
An isocurvature perturbation corresponds to a primordial (in the 
radiation era) perturbation of the matter which does not affect the geometry: it 
implies a perturbation in the relative composition of the cosmological fluid
with no perturbation in the {\it total} energy density. On the contrary 
an adiabatic perturbation is a primordial perturbation in the total 
energy density without modification of the relative quantities of the fluids.
For an adiabatic perturbation,  the relation between the 
matter and radiation density perturbations is by definition $\d_\g=(4/3)\d_m$,
which implies, during matter domination,
\beq
\d_\g\simeq {4\over 3}\d_T.\label{adi}
\eeq
  As will be seen later, the consequence of this 
relation together with the Poisson equation (\ref{poisson}) 
is that, both for the flat and open cases, the intrinsic 
fluctuations resulting from  adiabatic initial perturbations
{\it on scales larger than the Hubble radius} are always negligible. 
For an isocurvature perturbation, during matter domination,
\beq
\d_\g\simeq -{4\over 3}S,\label{iso}
\eeq
where $S=\d_m-(3/4)\d_\g$ is the entropy perturbation. Both for open and
flat universes, the appropriate combination of the equations of motion
(i.e. conservation and Euler equations) of the two fluids, radiation and 
baryons, tells us that $S$ is constant on scales bigger than the Hubble radius,
which is the case for perturbations we are interested in.  
Isocurvature perturbations also produce a gravitational potential. Using
the conservation and Euler equations for the two fluids, one obtains 
that the gravitational potential  is given, at the beginning of the matter era,
by
\beq
\Phi=-{1\over 5}S,\label{isophi}
\eeq
for the modes larger than the Hubble radius.

To summarize, the multipole coefficients of  (\ref{fluct}) 
can be decomposed into the sum of contributions 
\beq
a_{lm}=a_{lm}^{int}+a_{lm}^{SW}=a_{lm}^{int}+a_{lm}^{pSW}+a_{lm}^{Dop}
+a_{lm}^{ISW}.
\eeq
The rest of the paper will consist in evaluating the various terms of this 
expression for open/flat universes with adiabatic/isocurvature perturbations.

\section{Flat universe}
In this section, we consider the flat FLRW models.
 To study the CMBR fluctuations, it is more convenient to work in
spherical coordinates with the observer at the center, naturally adapted
to observations on the celestial sphere, rather than
 the more usual cartesian coordinates:
\beq
dl^2=dr^2+r^2\left(d\theta^2+\sin^2\theta d\phi^2\right).
\eeq
Any perturbation $f(r,\theta,\phi)$ can then be decomposed into
\beq
f(r,\theta,\phi,\eta)=\int dk \sum_{lm} f_{klm}(\eta)Q_{klm},
\eeq
where the mode functions $Q_{klm}$ are the eigenfunctions of the spatial
Laplacian with the eigenvalues $-k^2$ and are defined by
\beq
Q_{klm}=\sqrt{2\over\pi} kj_l(kr)Y_{lm}(\theta,\phi),
\eeq
where the $j_l$ are  the spherical Bessel functions and $Y_{lm}$ 
are the spherical harmonics.
The definition of the $Q_{klm}$ is such that they are normalized i.e.
\beq
\int r^2dr\, \sin\theta \, d\theta \, d\phi\,  Q_{klm}^* Q_{k'l'm'}=
\d(k-k')\d_{ll'}\d_{mm'}.\label{orthonormal}
\eeq
This decomposition is the equivalent of the Fourier decomposition
but in a system of spherical coordinates. This presentation is a good 
introduction to the open case where the spherical coordinates are the most
natural spatial cooordinates.
 
All the quantities introduced in the previous section can now be 
decomposed on the mode functions $Q_{klm}$. As announced before, we shall
focus essentially on the gravitational potential $\Phi$. The linearized 
Einstein equations yield a second order differential equation for the 
evolution of each mode $\Phi_{klm}$, which shows that the modes larger than the 
Hubble radius, i.e. $k<h$, are constant in time in a {\it flat universe} 
during the matter era. 
This result has several consequences. The first is that the ISW contribution 
vanishes. The second is that the relation (\ref{v}) simplifies to
\beq
hV=-{2\over 3}\Phi, \qquad (K=0).\label{vflat}
\eeq
Now, using the orthonormality property (\ref{orthonormal}) with equations (\ref{sw})
and (\ref{vflat}), the Sachs-Wolfe harmonic coefficients can be expressed
in the form
\beq
a_{lm}^{SW}=\sqrt{2\over\pi}\int_0^\infty dk \, k {\Phi_{klm}\over 3}
{\cal W}_l^{SW}(k),
\eeq
with the window function ${\cal W}_l^{SW}(k)={\cal W}_l^{pSW}(k)
+{\cal W}_l^{Dop}(k)$, the proper SW and Doppler contributions being
respectively given by
\beq
{\cal W}_l^{pSW}(k)=j_l(kr_{ls}), \quad {\cal W}_l^{Dop}(k)=2h_{ls}^{-1}{d\over dr}j_l(kr_{ls})- 
2h_0^{-1}{d\over dr}j_l(kr_0).
\eeq
$r_{ls}$ is the comoving distance from the hypersurface $\eta=\eta_{ls}$. It is related to the
redshift $z_{ls}$ ($\equiv (a_0/a_{ls}) -1$) by the relation
\beq
r_{ls}=2h_0^{-1}\left(1-(1+z_{ls})^{-1/2}\right)=2\left(h_0^{-1}- 
h_{ls}^{-1}\right).
\eeq
Using the relations 
\beq
{d\over dr}j_l(kr)=kj_{l-1}(kr)-{l+1\over r}j_l(kr),\quad
(2l+1)j_l(x)=x(j_{l+1}+j_{l-1}(x)),
\eeq
one can rewrite the dipole term
\beq
{\cal W}_1^{SW}(k)={2\over 3}kh_0^{-1}\left(j_0(kr_{ls})-1\right)+
{kr_{ls}\over 3}(1-4h_{ls}^{-1}r_{ls}^{-1})j_2(kr_{ls}).
\eeq
and the terms $l>1$
\beq
{\cal W}_l^{SW}(k)=\left(1+{2l\over h_{ls}r_{ls}}\right)j_l(kr_{ls})
-{2k\over h_{ls}}j_{l+1}(kr_{ls}).
\eeq
Since $j_0(x)=\sin x/x$ and $j_l(x)\sim x^l$ for $x<<1$, the 
dipole term  ${\cal W}_1^{SW}(k)$ behaves like $(kr_{ls})^3$
for small $k$. On the other hand, for $l>1$, ${\cal W}_l^{SW}(k)$ goes
like $(kr_{ls})^l$ for long wavelengths. This means that the Sachs-Wolfe dipole on 
ultra large scales is suppressed with respect to the quadrupole.
Let us now consider the intrinsic contribution. In the case of adiabatic
perturbations, equation (\ref{adi}) and  the Poisson equation (\ref{poisson})
(with $K=0$) shows that the intrinsic contribution is always negligible with respect to the Sachs-Wolfe contribution
for modes larger than the Hubble radius at the time of last scattering.
For adiabatic perturbations, the temperature fluctuations reduce to the 
Sachs-Wolfe fluctuations. The (total) cosmological dipole 
is thus suppressed with respect to the quadrupole (this remarkable property was
stressed in \cite{bl}). If this case describes the 
reality we live in, the conclusion is that the observed dipole is 
necessarily of local origin.
 
For isocurvature perturbations, the intrinsic contribution cannot be neglected
and the total window function ${\cal W}_l(k)$ is the sum of 
the SW window function and of  the intrinsic window function 
which, using (\ref{decompos1}) and (\ref{iso}), as well as (\ref{isophi}),
 is found to be 
\beq
{\cal W}_l^{iso}(k)=5j_l(kr_{ls}).
\eeq
Hence ${\cal W}_1$ behaves like $kr_{ls}$ for small $k$ and can be made as large 
as one wishes with respect to the quadrupole window function 
by considering large enough scales. 

The above analysis has been dealing with  individual modes. To get the total 
multipoles, one must sum on all the modes.
Assuming that $\Phi$ is a gaussian random field described by the power spectrum
\beq
\langle\Phi_{klm}\Phi_{k'l'm'}\rangle
=2\pi^2k^{-3}{\cal P}_\Phi(k) \d(k-k')\d_{ll'}
\d_{mm'} \label{ps}
\eeq
 the expectation value of the various multipoles reads
\beq
\langle |a_{lm}|^2\rangle={4\pi\over 9}\int {dk\over k}{\cal P}_\Phi(k)
{\cal W}_l^2(k).
\eeq
Once the power spectrum is given, the difference between the 
multipoles comes only from the window functions. The most common power 
spectrum is the flat spectrum corresponding to a constant 
${\cal P}_\Phi(k)$.
\begin{figure}
\centering
\epsfig{figure=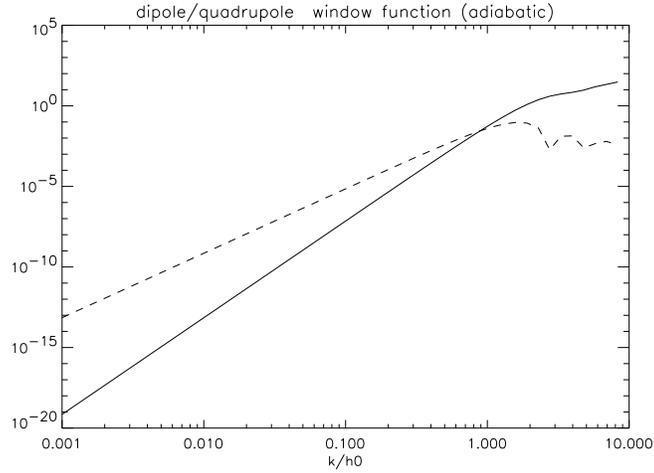, width=9cm}
\caption{squared amplitude of the window functions in the adiabatic case 
for the dipole (solid curve) and the quadrupole (dashed curve). On ultra
large scales the dipole window function is suppressed with respect to
the quadrupole one.}
\end{figure}
On Fig. 1, we have plotted the square of the dipole and quadrupole 
window functions in the adiabatic case. Since the dipole window function is 
always subdominant with respect to the quadrupole one or at best of the same  
order of magnitude for $k\sim h_0$, the conclusion is that the observed
dipole can only be explained by a local effect in that case. 
\begin{figure}
\centering
\epsfig{figure=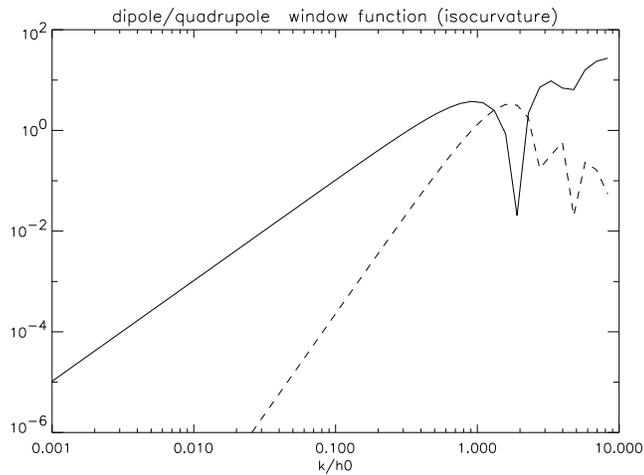, width=9cm}
\caption{squared amplitude of the window functions in the isocurvature case 
for the dipole (solid curve) and the quadrupole (dashed curve). On ultra
large scales the quadrupole window function is suppressed with respect to
the dipole one.}
\end{figure}
On the contrary, for isocurvature perturbations, Fig.2 shows that the 
dipole window function dominates on ultra large scales and the observed dipole 
can be of cosmological origin. However the corresponding isocurvature 
spectrum cannot be constant on all scales because the main contribution to 
the dipole and quadrupole would come from scales $k\sim h_0$ and the 
resulting dipole and quadrupole would be comparable.
To obtain cosmological dipole and quadrupole compatible with the observations, 
two conditions must therefore be satisfied: the existence of ultra large 
scale isocurvature perturbations; and their spectrum must be suppressed 
for scales $\lambda < 100 H_0^{-1}$.
Although this cut-off condition can appear on first thought difficult to fulfill in a sensible 
model, it turns out that two relatively simple ideas can produce such 
an effect. One idea is the existence of pre-inflationnary isocurvature 
perturbations, which would have been pushed away far beyond the horizon during
the inflation era \cite{turner}; the second idea is to consider the 
simplest two scalar field inflation model, which naturally  produces 
an isocurvature spectrum with an abrupt cut-off \cite{l96}.

\section{Open universe}

We now write the background spatial metric  in the form
\beq
dl^2=dr^2+\sinh^2r(d\theta^2+\sin^2\theta d\phi^2).
\eeq
With this choice of coordinates the Friedmann equation reads
\beq
H^2={8\pi G\over 3}\rho+{1\over a^2},
\eeq
where $H=ah$ is the Hubble parameter. As is clear from the above formula, 
$a$ represents the curvature scale: on scales smaller than $a$, the curvature
of space is not ``felt'' whereas scales larger than $a$ are 
affected by the curvature. 
It is usual to parametrize the space curvature by the ratio of the 
energy density with respect to the critical energy density today, 
$\Omega_0=3H_0^2/8\pi G$. $\Omega_0=1$ for a flat space and $\Omega_0 <1$ 
for an open space. It then follows, in the matter era, that  the 
comoving Hubble parameter can be expressed in terms of the redshift $z$
and $\Omega_0$:
\beq
h=\sqrt{1+\Omega_0 z\over 1-\Omega_0}.\label{h}
\eeq  
In particular $h_0=1/\sqrt{1-\Omega_0}$ represents the ratio between the 
current curvature scale $a_0$ and the Hubble scale $H_0^{-1}$. For an 
open universe the curvature scale is thus always larger than the Hubble
scale and all the more larger as $\Omega_0$ is closer to one.

In an open universe, the Fourier treatment does not apply. One must then 
use the formalism of mode functions for an hyperbolic space.
Following \cite{lw}, any gaussian random field
can be expanded in the form 
\beq
f(r,\theta,\phi,\eta)=\int_0^\infty dk \sum_{l=0}^\infty\sum_{m=-l}^{l}
 f_{klm}(\eta) Q_{klm},\label{basis}
\eeq
where the $Q_{klm}$ are eigenfunctions of the spatial 
Laplacian with eigenvalues $-k^2$. They can be chosen of the form 
\beq
Q_{klm}=\Pi_{kl}(r) Y_{lm}(\theta,\phi),
\eeq
where  $Y_{lm}(\theta,\phi)$ are the usual spherical harmonics on the 
two-sphere. The radial mode functions $\Pi_{kl}$ can be classified into
two categories depending on the sign of $q^2\equiv k^2-1$.
The radial functions for the subcurvature modes, corresponding to the 
values $q^2>0$, are given by 
\beq
\Pi_{kl}\equiv N_{kl}\tilde\Pi_{kl},
\eeq
with
\beq
N_{kl}=\sqrt{2\over\pi} q^2\left[\prod_{n=0}^l (n^2+q^2)\right]^{-1/2},
\quad \tilde\Pi_{kl}=q^{-2}(\sinh r)^l\left[{-1\over\sinh r}{d\over dr}\right]
^{l+1}\cos (qr).
\eeq
The subcurvature modes provide a complete orthonormal basis for 
square integrable functions and for this reason were the only modes 
taken into account by cosmologists. However, as stressed recently by Lyth 
and Woszczyna \cite{lw}, they are not enough to describe properly gaussian 
random fields and  one must also include the supercurvature modes, corresponding to the 
values $-1<q^2<0$. Their radial functions  are obtained from the subcurvature
 modes by analytic continuation: 
\beq
\Pi_{kl}\equiv N_{kl}\tilde\Pi_{kl},
\eeq
with
\beq
N_{k0}=\sqrt{2\over\pi} |q|, \qquad
N_{kl}=\sqrt{2\over\pi} |q|\left[\prod_{n=1}^l (n^2+q^2)\right]^{-1/2}
\quad (l>0)
\eeq
and
\beq
\tilde\Pi_{kl}=|q|^{-2}(\sinh r)^l\left[{-1\over\sinh r}
{d\over dr}\right]^{l+1}\cosh (|q|r).
\eeq
The normalization of the mode functions is such that 
\beq
\int_0^\infty \Pi_{kl}(r)\Pi_{k'l'}(r)\sinh^2r dr=\d(q-q')\d_{ll'}.
\eeq
Moreover for a homogeneous gaussian random field, one defines the power 
spectrum by
\beq
\langle f_{klm}f^*_{k'l'm'}\rangle={2\pi^2\over k|q|^2}{\cal P}_f(k)\d(k-k')
\d_{ll'}\d_{mm'}.
\eeq
Note that in the flat space limit, corresponding to the limits 
$k\rightarrow \infty$ and $r\rightarrow 0$ with $kr$ fixed, 
\beq
\Pi_{kl}(r)\rightarrow \sqrt{2\over\pi} kj_l(kr),
\eeq
and one  recovers equation (\ref{ps}) for the power spectrum.

In contrast with the flat case, the gravitational potential modes
$\Phi_{klm}$ on scales larger than the Hubble radius evolve with time. Their
evolution is given by the linearized Einstein equations (see e.g. \cite{mfb}):
\beq
\Phi_k(\eta)=F(\eta){\tilde \Phi}_k, 
\eeq
with 
\beq
F(\eta)=5{\sinh^2\eta-3\eta \sinh\eta+4\cosh\eta-4\over (\cosh\eta-1)^3}.
\label{f}   
\eeq
(with the normalization $a=\cosh\eta -1$ for
the scale factor). F is normalized so that $F\rightarrow 1$ when $\eta\rightarrow 0$. 
${\tilde \Phi}_k$ is constant in time and is given by the initial spectrum
at the beginning of the matter era. At that initial stage, the curvature can
 be ignored. In fact,  even for small values 
of $\Omega_0$,  $F(\eta)$ remains almost constant during most of the matter 
era and drops suddenly in the recent past. Similarly, the velocity potential
obeys to a law different from the flat case and given by
\beq
hV(\eta)=-{2\over 3}G(\eta)\tilde\Phi,
\eeq
with
\beq
 \quad G(\eta)= {{15\,
  \left( 2\,\eta + \eta\,\cosh\eta - 3\,\sinh\eta \right) \,\sinh\eta}
     \over {16 \sinh^6(\eta/2)}}.
\eeq
Like $F$, $G$ has been normalized such that $G\rightarrow 1$ when $\eta\rightarrow
0$ and $G$ is significantly  different from one only for small redshifts.

The decomposition of the Sachs-Wolfe terms in (\ref{sw}) on the basis
$Q_{klm}$, like in (\ref{basis}), yields for the harmonic coefficients
 the expression 
\beq
a_{lm}=\sqrt{2\over \pi}\int_0^\infty  dk\, |q| {1\over 3}\tilde\Phi_{klm}
\left[{\cal W}_{kl}^{pSW}+{\cal W}_{kl}^{Dop}+{\cal W}_{kl}^{ISW}\right],
\eeq
with
\beq
{\cal W}_{kl}^{pSW}=(3F_{ls}-2G_{ls})\hat\Pi_{kl}(r_{ls})/|q|,\label{win1}
\eeq
\beq
{\cal W}_{kl}^{Dop}=2h_{ls}^{-1}G_{ls}\partial_r\hat\Pi_{kl}(r_{\rm ls})/|q|
-{2\over 3}h_0^{-1}G_0 \,  k\d_{l,1}, \label{win2}
\eeq
and 
\beq
{\cal W}_{kl}^{ISW}= \left[-6F_{ls}\hat\Pi_{kl}(r_{\rm ls})
+6\int_0^{r_{\rm ls}} dr\, F(\eta_{\rm ls}+r_{\rm ls}-r)\partial_r\hat\Pi_{kl}(r)\right]/|q|.
\label{win3}
\eeq
Except for the flat case limit, we shall take $F_{ls}=G_{ls}=1$
in the above expressions, which is an excellent approximation because
$z_{ls}>>1$. A hat means a division
by $\sqrt{2/\pi}$.
$h_{\rm ls}$ is the comoving Hubble parameter at the time of last scattering. 
$r_{\rm ls}$ is the radius coordinate corresponding to the emission of the photons,
i.e. the coordinate of the last scattering surface. This radius can be 
related to the redshift $z_{\rm ls}$ via the relation
\beq
\sinh r(z)={2\over
h_0}(1+z)^{-1}\Omega_0^{-2}\left[\Omega_0z+(\Omega_0-2)
\left((1+\Omega_0 z)^{1/2}-1\right)\right].\label{ropen}
\eeq

We must also consider the intrinsic contribution $a_{lm}^{int}$ to the 
harmonic coefficients. For adiabatic fluctuations, like in the flat case,
the intrinsic contribution is negligible as can be seen from (\ref{poisson}): 
the reason is not this time
the smallness of $k^2$ but the fact that $h^2>>1$ (see  (\ref{h})). 
For isocurvature perturbations, the intrinsic window function  is given by
\beq
{\cal W}_{kl}^{iso}=5F_{ls} \hat\Pi_{kl}(r_{\rm ls})/|q|,
\eeq

Finally, once a power spectrum is given, the expectation values for the 
harmonic coefficients can then be written in the form 
\beq
\langle |a_{lm}|^2\rangle={4\pi\over 9}\int {dk\over k}{\cal P}_\Phi(k)
{\cal W}_l^2(k),
\eeq
where ${\cal W}_l(k)$ is the sum of all the relevant window functions.

\subsection{Ultra large scale limit}
In the flat case, it was possible to obtain a dipole much larger than 
the quadrupole with isocurvature perturbations on scales much larger than 
the Hubble radius. We wish now to consider such scales in an open universe.
In the ultra large scale limit, corresponding to $k\rightarrow 0$, the 
general expressions can be simplified because $\Pi_{kl}(r)\sim k f_l(r)$.
Following \cite{gllw}, we define 
\beq
N_l\equiv \lim_{k\rightarrow 0} kN_{kl}=\sqrt{2\over\pi}\prod_{n=2}^l
(n^2-1)^{-1/2}
\eeq
for $l>1$ and $N_1=\sqrt{2/\pi}$. Similarly, we define
\beq
\tilde\Pi_l\equiv\lim_{k\rightarrow 0}\tilde\Pi_{kl}/k^2.
\eeq
The expressions we are mainly interested in are 
\beq
\tilde\Pi_1(r)={1\over 2}\left(\coth r- {r\over\sinh^2 r}\right)
\eeq
and
\beq
\tilde\Pi_2(r)= {1\over 2}\left[1+3{1-r\coth r\over\sinh^2r}\right].
\eeq
The window functions (\ref{win1}-\ref{win3}) can then be expressed in the limit $k\rightarrow 0$
in the form
\beq
{\cal W}_{kl}\sim k {\cal L}_l (\Omega_0, z_{\rm ls}),
\eeq
with 
\beq
{\cal L}_l^{pSW}=\hat N_l\tilde\Pi_l(r_{\rm ls}), \quad 
{\cal L}_l^{Dop}=2\hat N_l h_{\rm ls}^{-1}{d\over dr}\tilde\Pi_l(r_{\rm ls})
-{2\over 3}h_0^{-1}G_0\d_{l,1},
\eeq
and
\beq
{\cal L}_l^{ISW}=\hat N_l\left[-6 \tilde\Pi_l(r_{\rm ls})
+6\int_0^{r_{\rm ls}}dr\,  F(\eta_{\rm ls}+r_{\rm ls}-r)
{d\over dr}\tilde\Pi_l(r)\right].
\eeq
\begin{figure}
\centering
\epsfig{figure=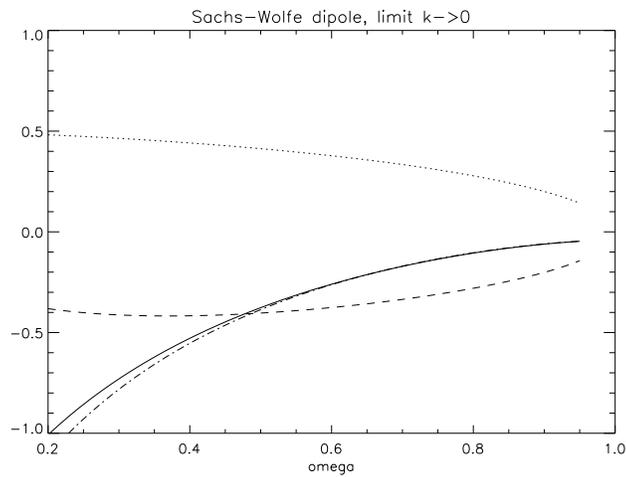, width=9cm}
\caption{amplitude of the Sachs Wolfe dipole window functions in the limit 
$k\rightarrow 0$
as a function of $\Omega_0$. The total Sachs-Wolfe window function 
(solid curve) is the sum of the pure Sachs-Wolfe contribution (dotted curve), 
of the Doppler contribution (dashed curve), which almost cancel each other,
and finally of the ISW contribution (dashed-dotted curve).}
\end{figure}
\begin{figure}
\centering
\epsfig{figure=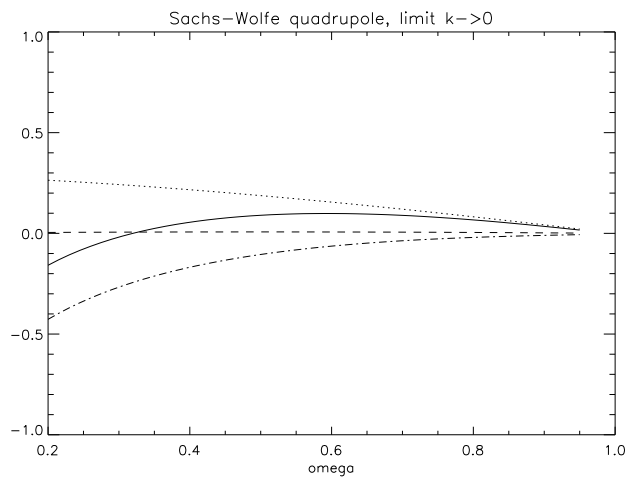, width=9cm}
\caption{amplitude of the Sachs Wolfe quadrupole window functions in the limit 
$k\rightarrow 0$ as a function of $\Omega_0$. In the total Sachs-Wolfe window function
(solid curve), The Doppler term (dashed curve) is very 
small with respect to the two other terms, the pure SW term (dotted curve)
and the ISW term (dashed-dotted curve). One notices that the SW dipole
vanishes near $\Omega_0\simeq 0.3$.}
\end{figure}
These three quantities, as well as their sum corresponding to the total 
Sachs-Wolfe effect are plotted as functions of $\Omega_0$ for the dipole
(Fig. 3) and for the quadrupole (Fig. 4). The first striking difference 
with the flat case is that the Sachs-Wolfe dipole is not suppressed 
with respect to the quadrupole. One can see more precisely on Fig. 3
that this is due principally to the ISW contribution, even if the pure SW
 and   Doppler contributions do not cancel each other exactly, especially 
for low $\Omega_0$. As for the quadrupole, one can see that the Doppler 
contribution is extremely small in comparison with the other terms. 
Another remark is that the quadrupole vanishes around $\Omega_0\simeq 0.3$
for which the pure SW effect and the ISW effect happen to compensate 
each other. 

\begin{figure}
\centering
\epsfig{figure=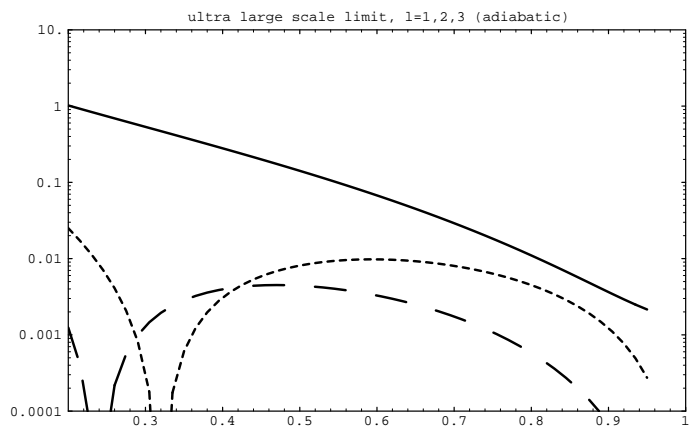, width=9cm}
\caption{squared amplitude of the total window functions (intrinsic and Sachs-Wolfe) 
for ultra large scale adiabatic perturbations (limit $k\rightarrow 0$), 
for the three multipoles $l=1,2,3$ (from top to bottom). }
\end{figure}
\begin{figure}
\centering
\epsfig{figure=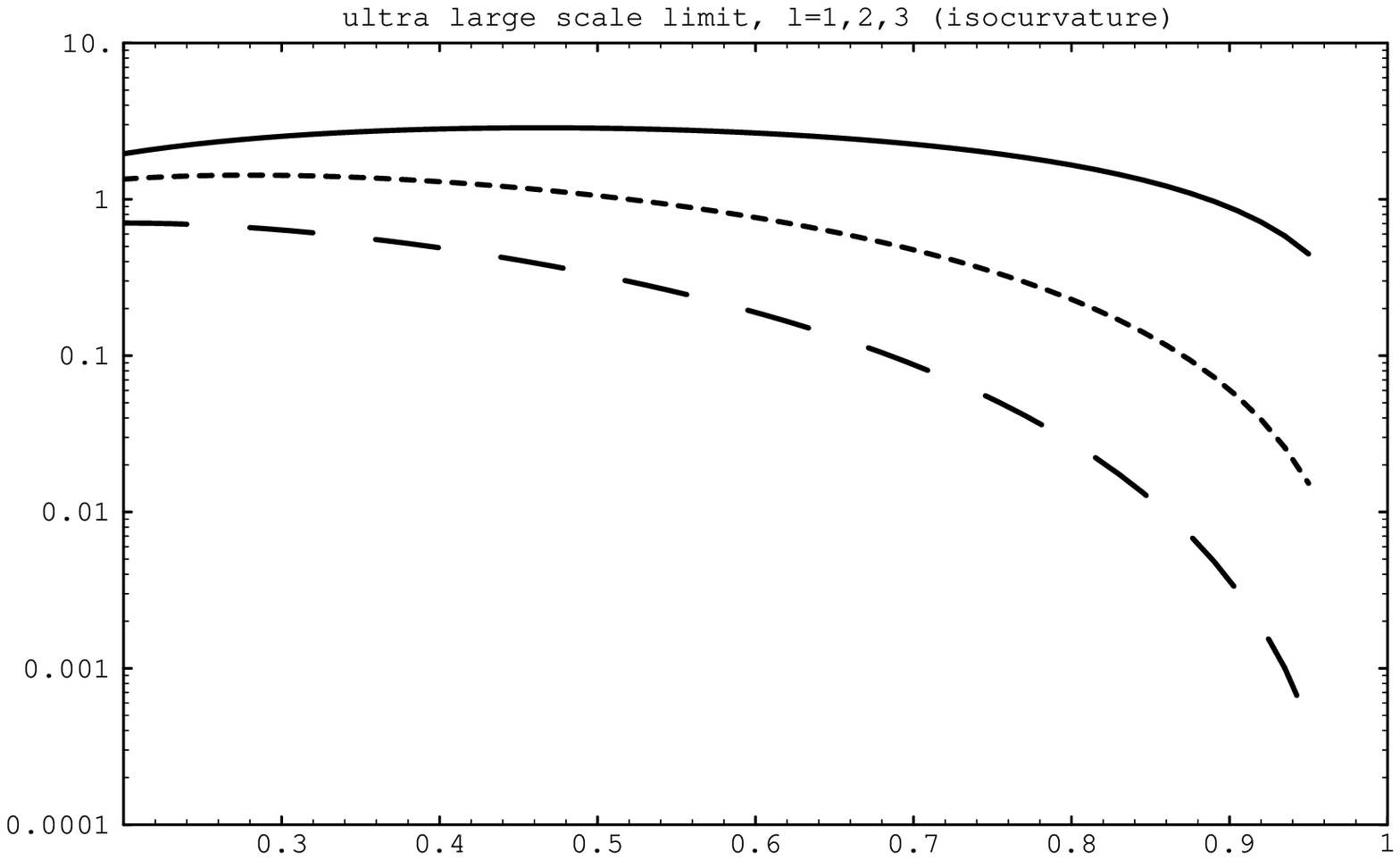, width=9cm}
\caption{squared amplitude of the total window functions (intrinsic and Sachs-Wolfe) 
for ultra large scale isocurvature perturbations (limit $k\rightarrow 0$), 
for the three multipoles $l=1,2,3$ (from top to bottom). }
\end{figure}
In Fig. 5 and Fig. 6, we have plotted, still in the ultra large scale limit, 
the total window functions for the multipoles $l=1,2,3$ with, respectively,
 adiabatic and isocurvature initial conditions. We have added the multipole $l=3$ to
show that even if the dipole can be made much bigger than the quadrupole
in the zone $\Omega_0\simeq 0.3$, this cannot explain the observed CMBR since the 
next multipole is roughly one tenth of  the dipole.
The conclusion is therefore that, in an {\it open} universe, the {\it 
cosmological dipole} is typically of the {\it same order of magnitude}
as the higher multipoles. Within our assumptions, the observed dipole
can be explained cosmologically only in a flat space. In the next section,
we examine in detail this flat space limit in order to put a limit
on the flatness required to explain the dipole by cosmological perturbations.

\subsection{Flat space limit}
We saw in the previous subsection that, for supercurvature modes, the 
window functions go to zero when $\Omega\rightarrow 1$. We need to examine
now in more details what is the relative importance of 
the dipole with respect to the quadrupole in this limit. To do so, it is 
convenient to introduce the paramater
\beq
\epsilon=\sqrt{1-\Omega_0}.
\eeq
One can immediately notice that the curvature radius reads $a_0=H_0^{-1}/
\epsilon$ so that the curvature radius goes to infinity with respect to 
the Hubble radius, as one can expect in a flat space limit.
Expanding (\ref{ropen}) in $\e$, one  finds
\begin{eqnarray}
r(\Omega_0,z)&=&2\left(1-{1\over \sqrt{1+z}}\right)\epsilon +\left[{2\over 1+z}
\left(z+3-3\sqrt{1+z}+{z\over 2\sqrt{1+z}}\right) \right. \nonumber\\
& &\left.  -{4\over 3}\left(1-{1\over\sqrt{1+z}}
\right)^3\right]\e^3+{\cal O}(\e^5).
\end{eqnarray}
The expansion of the terms appearing in the total Sachs-Wolfe effect is 
\begin{eqnarray}
{\cal W}_{k1}^{pSW}&=&{2\over 3}k\left(1-{1\over \sqrt{1+z}}\right)\epsilon 
+{2\over 3}k\left[(z+{11\over 7}-{22+35z\over 14\sqrt{1+z}}
)/(1+z) \right. \nonumber \\
& & \left.
-{2\over 5}(3+k^2)\left(1-{1\over\sqrt{1+z}}\right)^3\right]\epsilon^3
+{\cal O}(\e^5),\label{d1}
\end{eqnarray}
\begin{eqnarray}
{\cal W}_{k1}^{Dop}&=&-{2\over 3}k\left(1-{1\over \sqrt{1+z}}\right)\epsilon
+{2\over 3}k\left[{1\over 7}+{1\over\sqrt{1+z}}\left({7z-2\over 14(1+z)}
\right. \right. \nonumber \\
& &\left. \left.
-{2\over 5}(4+3k^2)\left(1-{1\over \sqrt{1+z}}\right)^2\right)
\right]\e^3 +{\cal O}(\e^5),
\end{eqnarray}
\beq
{\cal W}_{k1}^{ISW}=-{16k\over 21}\left(1-{1\over \sqrt{1+z}}\right)^2
\left(1+{2\over\sqrt{1+z}}\right)\e^3+{\cal O}(\e^5).\label{d3}
\eeq
Note that the strict limit $\Omega_0=1$ is not obtained simply by setting
$\epsilon=0$, which would give zero for all the terms above. Indeed it makes
more sense to label the perturbation scales by the physical quantity 
$k/h_0$ instead of the unphysical wavenumber $k$. Since $h_0=1/\epsilon$, the 
proper flat limit is given by $\e\rightarrow 0$ with $k\e$ fixed. However, 
this limit applies only to subcurvature modes whereas supercurvature modes
correspond to larger and larger scales as $\Omega_0\rightarrow 1$ since 
they are bounded by $k<1$. The most striking feature of (\ref{d1}-\ref{d3})
is that 
the dominant terms of the pure SW contribution and of the Doppler term 
just cancel each other. It is thus clear that the suppression of the 
dipole is really a consequence of the flatness.
 We have also written explicitly the terms at the order 
$\e^3$ because they represent the leading order in the total SW window
function. 
For the quadrupole, one obtains
\beq
{\cal W}_{k2}^{pSW}={4\over 15}k(k^2+3)^{1/2}
\left(1-{1\over \sqrt{1+z}}\right)^2\epsilon^2 +{\cal O}(\e^4),\label{q1}
\eeq
\beq
 {\cal W}_{k2}^{Dop}={8\over 15}k(k^2+3)^{1/2}
\left({1\over \sqrt{1+z}}-{1\over 1+z}
\right)\epsilon^2 +{\cal O}(\e^4),
\eeq
\beq
{\cal W}_{k1}^{ISW}={\cal O}(\e^4).\label{q3}
\eeq
For isocurvature, one must add the intrinsic contribution which is roughly 
five times ${\cal W}_{kl}^{pSW}$.
By comparing (\ref{d1}-\ref{d3}) with (\ref{q1}-\ref{q3}), one can see that
for $k<<1$ all the window functions, both for the dipole and the quadrupole,
are proportional to $k$. As a consequence one cannot obtain a significant 
dipole by considering sufficiently large scales, like in the flat case.
The only possible difference between the dipole and the quadrupole can thus  
come only from the parameter $\e$. For adiabatic perturbations, the 
cancellation of the $\e$ terms implies that the dipole is $\e$ times the
quadrupole. On the contrary, for isocurvature perturbations, the total 
dipole is proportional to $\e$ and can be much bigger than the quadrupole 
which goes like $\e^2$. To obtain a dipole $10^2$ times larger than the 
quadrupole, one thus needs $\e < 10^{-2}$, i.e. $|\Omega_0 -1|<10^{-4}$.

\section{Conclusions}
The main conclusion of this paper is that a cosmological origin for 
the dipole (or a significant part of it) must be rejected for an open 
cosmology, which means  that the observed dipole would be in that case
 essentially a Doppler effect resulting from our local motion dominated by
 small scale density perturbations.
 This conclusion assumes that the universe can be described as a FLRW 
model with  gaussian fields of linear perturbations.
 More exotic possibilities have not been considered here. 

Another conclusion of this work is that the suppression in a flat universe
 of the dipole due to ultra large scale adiabatic perturbations,
with respect to the quadrupole,
is no longer true in an open universe.

It turns out that the only possibility to get a dipole two orders of
magnitude bigger than the other multipoles is a model with isocurvature 
perturbations on scales of the order of $10^2 H_0^{-1}$ in an almost 
flat geometry:
the constraint is $|\Omega_0 -1|<10^{-4}$. This constraint can also 
be interpreted as demanding that the curvature scale be one hundred timed
larger than the Hubble radius. When this constraint is satisfied but 
with $\Omega_0$ non strictly equal to one (this would be the case 
for example of an open universe after a phase of inflation), two kinds
of perturbations can produce the observed dipole : the subcurvature modes
on scales larger than one hundred times the Hubble radius, like in the 
flat case; but also the supercurvature modes, which do not exist in the 
flat case.

To conclude, a reliable  measurement of $\Omega_0$ (see \cite{dbw} 
for a discussion on measuring $\Omega_0$) different from  one
would strongly confirm the Doppler assumption for the dipole. 
Conversely, if unquestionable  cosmological observations conclude that our 
local motion is in disagreement with the CMBR dipole, then it would 
suggest that we live in a flat universe and that ultra large scale 
isocurvature perturbations must exist.

\bigskip
\bigskip\noindent
{\bf Acknowledgement.}
\medskip

I would like to thank N. Deruelle, T. Piran and D. Wands for stimulating 
discussions.

\end{document}